\documentclass[a4paper,11pt]{article}

\usepackage{url}
\usepackage{mathtools}
\usepackage{amssymb}
\usepackage{graphicx,caption}
\usepackage{a4wide}

\newcommand{\cc}{\ensuremath{\mbox{\it cc}}}
\newcommand{\tr}{\ensuremath{\mbox{\it tr}}}
\newcommand{\rc}{\ensuremath{\mbox{\it rc}}}
\newcommand{\weight}{\ensuremath{\omega}}
\newcommand{\demidelta}{\frac{\Delta}{2}}
\newcommand*{\diff}{\mathop{}\!\mathrm{d}}

\date{}

\title{{\bf Weighted, Bipartite, or Directed Stream Graphs\\\smallskip for the Modeling of Temporal Networks}}

\author{Matthieu Latapy\,\footnote{Corresponding author: {Matthieu.Latapy@lip6.fr}} ,\ \  Cl\'emence Magnien,\ \  Tiphaine Viard
\smallskip\\
{\small Sorbonne Universit\'e, CNRS, LIP6, F-75005 Paris, France}}

\begin{document}

\maketitle

\begin{abstract}
{
We recently introduced a formalism for the modeling of temporal networks, that we call stream graphs. It emphasizes the streaming nature of data and allows rigorous definitions of many important concepts generalizing classical graphs. This includes in particular size, density, clique, neighborhood, degree, clustering coefficient, and transitivity. In this contribution, we show that, like graphs, stream graphs may be extended to cope with bipartite structures, with node and link weights, or with link directions. We review the main bipartite, weighted or directed graph concepts proposed in the literature, we generalize them to the cases of bipartite, weighted, or directed stream graphs, and we show that obtained concepts are consistent with graph and stream graph ones. This provides a formal ground for an accurate modeling of the many temporal networks that have one or several of these features.
}
\end{abstract}

\section{Introduction}
\label{sec:introduction}

Graph theory is one of the main formalisms behind network science. It provides concepts and methods for the study of networks, and it is fueled by questions and challenges raised by them. Its core principle is to model networks as sets of nodes and links between them. Then, a graph $G$ is defined by a set of nodes $V$ and a set of links $E \subseteq V \otimes V$ where each link is an undordered pair of nodes\,\footnote{Given any two sets $X$ and $Y$, we denote by $X\times Y$ the cartesian product of $X$ and $Y$, {\em i.e.} the set of all ordered pairs $(x,y)$ such that $x \in X$ and $y \in Y$. We denote by $X \otimes Y$ the set of all unordered pairs composed of $x\in X$ and $y\in Y$, with $x\neq y$, that we denote by $xy=yx$.}. In many cases, though, this does not capture key features of the modeled network. In particular, links may be weighted or directed, nodes may be of different kinds, etc. One key strength of graph theory is that is easily copes with such situations by defining natural extensions of basic graphs, typically weighted, bipartite, or directed graphs. Classical concepts on graphs are then extended to these more complex cases.

\begin{figure}[!h]
\centering
\includegraphics[scale=.5]{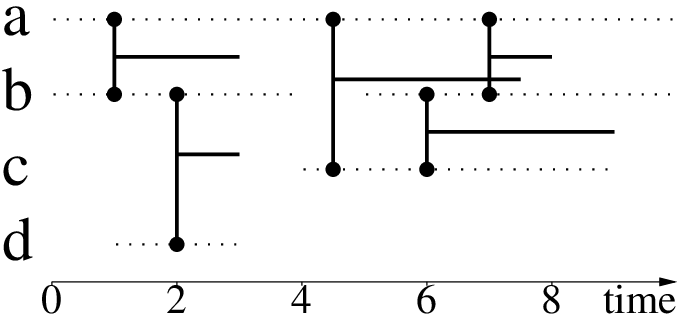}
\caption{
{\bf An example of stream graph:}
$S=(T,V,W,E)$ with $T = [0,10] \subseteq \mathbb{R}$, $V = \{a,b,c,d\}$, $W = [0,10]\times\{a\} \cup ([0,4]\cup[5,10])\times\{b\} \cup [4,9]\times\{c\} \cup [1,3]\times\{d\}$, and $E = ([1,3]\cup[7,8])\times\{ab\} \cup [4.5,7.5]\times\{ac\} \cup [6,9]\times\{bc\} \cup [2,3]\times\{bd\}$. In other words, $T_a = [0,10]$, $T_b = [0,4]\cup[5,10]$, $T_c = [4,9]$, $T_d = [1,3]$, $T_{ab} = [1,3]\cup[7,8]$, $T_{ac} = [4.5,7.5]$, $T_{bc} = [6,9]$, $T_{bd} = [2,3]$, and $T_{ad} = T_{cd} = \emptyset$.
}
\label{fig:ex}
\end{figure}

Stream graphs were recently introduced as a formal framework for temporal networks \cite{stream}, similar to what graph theory is to networks.
A stream graph $S$ is defined by a time set $T$, a node set $V$, a set of temporal nodes $W \subseteq T\times V$ and a set of temporal links $E \subseteq T\times V\otimes V$. See Figure~\ref{fig:ex} for an illustration. Each node $v \in V$ has a set of presence times $T_v = \{t, (t,v) \in W\}$. Likewise, $T_{uv} = \{t, (t,uv)\in E\}$ is the set of presence times of link $uv$. Conversely, $V_t = \{v, (t,v)\in W\}$ and $E_t = \{uv, (t,uv)\in E\}$ are the set of nodes and links present at time $t$, leading to the graph at time $t$: $G_t=(V_t,E_t)$. The graph induced by $S$ is $G(S) = (\{v, T_v\neq\emptyset\},\{uv, T_{uv}\neq\emptyset\})$.

Stream graphs encode the same information as Time Varying Graphs (TVG) \cite{DBLP:journals/paapp/CasteigtsFQS12}, Relational Event Models (REM) \cite{SOME:SOME203,Stadtfeld2017}, Multi-Aspect Graphs (MAG) \cite{DBLP:journals/tcs/WehmuthFZ16,DBLP:conf/dsaa/WehmuthZF15}, or other models of temporal networks. Stream graphs emphasize the streaming nature of data, but all stream graph concepts may easily be translated to these other points of views.

A wide range of graph concepts have been extended to stream graphs \cite{stream}. The most basic ones are probably the number of nodes $n = \sum_{v \in V} \frac{|T_v|}{|T|}$ and the number of links $m = \sum_{uv \in V\otimes V} \frac{|T_{uv}|}{|T|}$. Then, the neighborhood of node $v$ is $N(v) = \{(t,u), (t,uv) \in E\}$ and its degree is $d(v) = \frac{|N(v)|}{|T|}$.
The average degree of $S$ is the average degree of all nodes weighted by their presence time:
$d(S) = \sum_{v\in V} \frac{|T_v|}{|W|} d(v)$.

Going further, the density of $S$ is $\delta(S)=\frac{m}{\sum_{uv \in V\otimes V}|T_u\cap T_v|}$. It is the probability, when one chooses at random a time instant and two nodes present at that time, that these two nodes are linked together at that time.
Then, a clique is a subset $C$ of $W$ such that for all $(t,u)$ and $(t,v)$ in $C$, $u$ and $v$ are linked together at time $t$ in $S$, {\em i.e.} $(t,uv)\in E$. Equivalently, a subset of $W$ is a clique of $S$ if the substream it induces has density $1$.

This leads to the definition of clustering coefficient in stream graphs: like in graphs, $\cc(v)$ is the density of the neighborhood of $v$. Equivalently, $\cc(v) = \sum_{uw\in V\otimes V}\frac{|T_{vu} \cap T_{vw} \cap T_{uw}|}{|T_{vu} \cap T_{vw}|}$.
Likewise, the transitivity of $S$ is the fraction of all 4-uplets $(t,u,v,w)$ with $(t,uv)$ and $(t,vw)$ in $E$ such that $(t,vw)$ is also in $E$.









These concepts generalize graph concepts in the following sense. A stream $S$ is called {\em graph-equivalent} if it has no dynamics: $G_t = G(S)$ for all $t$. In this case, each stream property of $S$ is equal to the corresponding graph property of $G(S)$. For instance, the density of $S$ is equal to the one of $G(S)$. Graphs may therefore be seen as special cases of stream graphs (the ones with no dynamics).

\begin{figure}[!h]
\centering
\includegraphics[width=.85\textwidth]{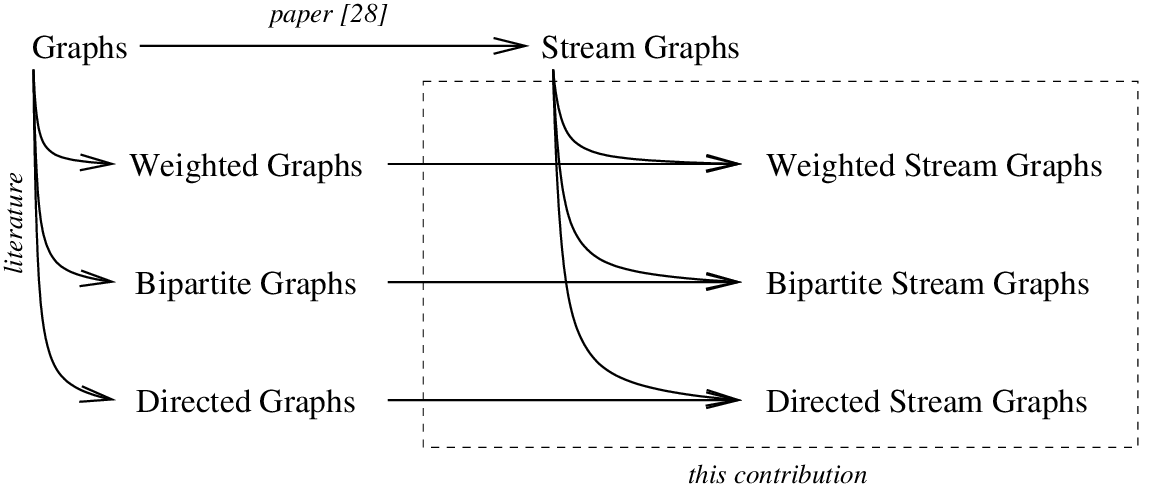}
\caption{
The global positionning of this contribution with respect to state-of-the-art. Left: weighted, bipartite and directed extensions of graph properties are available in the scientific literature. Top: a generalization of graphs to stream graphs was proposed in \cite{stream}. Dotted rectangle: in this contribution, we extend weighted, bipartite and directed graph concepts to weighted, bipartite, and directed stream graphs, as well as standard stream graph concepts to the weighted, bipartite and directed cases, in a consistent way.
}
\label{fig:schema}
\end{figure}

Like for graphs, the stream graph formalism was designed to be readily extendable to weighted, directed, or bipartite cases. However, these extensions remain to be done, and this is the goal of the present contribution, summarized in Figure~\ref{fig:schema}.

%
%
%
%
%
%

Before entering in the core of this contribution, notice that the set of available graph concepts is huge, much larger than what may be considered here. We therefore focus on a the set of key properties succintly summarized above. In particular, we do not consider path-related concepts, which would deserve a dedicated work of their own. Rather than being exhaustive, our aim is to illustrate how weighted, bipartite, or directed graph concepts may be generalized to stream graphs in a consistent way, and to provide a ground for further generalizations.

\section{Weighted stream graphs}
\label{sec:weighted}

A {\bf weighted graph} is a graph $G=(V,E)$ equipped with a weight function $\weight$ generally defined over $E$, and sometimes on $V$ too. Then, $\weight(v)$ is the weight of node $v$ and $\weight(uv)$ the one of link $uv$.
Link weights may represent tie strength (in a friendship or collaboration network for instance)~\cite{newman01scientific2, pnas09serrano},
link capacity (in an road or computer network, for instance)~\cite{po16candeloro, pnas09serrano},
or a level of similarity (in document or image networks, or gene networks for instance)~\cite{ac07kalna,sagmb05zhang}.
Node weights may represent reliability, availability, size, etc. As a consequence, weighted graphs are very important and they are used to model a wide variety of networks. In most cases, though, nodes are considered unweighted. We will therefore only consider weighted links in the following, except where specified otherwise.

Even when one considers a weighted graph $G=(V,E)$ equipped with the weight function $\weight$, the properties of $G$ itself (without weights) are of crucial interest. In addition, one may consider thresholded versions of $G$, defined as $G_{\tau} = (V_\tau,E_\tau)$ where $V_\tau = \{v \in V, \weight(v) \ge \tau\}$ and $E_\tau = \{uv \in E, \weight(uv)\ge\tau\}$, for various thresholds $\tau$. This actually is a widely used way to deal with graph weights, formalized in a systematic way as early as 1969 \cite{s69doreian}. However, one often needs to truly take weights into account, without removing any information. In particular, the importance of weak links is missed by thresholding approaches. In addition, determining appropriate thresholds is a challenge in itself \cite{cn18asfahlani,pnas09serrano,embs15smith}.

As a consequence, several extensions of classical graph concepts have been introduced to incoporate weight information and deal directly with it. The most basic ones are the maximal, minimal, and average weights, denoted respectively by $\weight_{\max}$, $\weight_{\min}$ and $\langle \weight\rangle$. The minimal weight $\weight_{\min}$ is often implicitely considered as equal to $0$, and weights are sometimes normalized in order to ensure that $\weight_{\max} = 1$ \cite{pre05onnela,sagmb05zhang,pre05grindrod,pre07ahnert}.

In addition to these trivial metrics, one of the most classical property probably is the weighted version of node degree, known as node strength \cite{pnas04barrat,pre04newman,ddns08antoniou}: $s(v) = \sum_{u \in N(v)} \weight(uv)$. Notice that the average node strength is equal to the product of the average node degree and the average link weight.

Strength is generally used jointly with classical degree, in particular to investigate correlations between degree and strength: if weights represent a kind of activity (like travels or communications) then correlations give information on how activity is distributed over the structure \cite{pnas04barrat,jasist09panzarasa}. One may also combine node degree and strength in order to obtain a measure of node importance. For instance, \cite{sn10opsahl} uses a tuning parameter $\alpha$ and compute $d(v)\cdot\left(\frac{s(v)}{d(v)}\right)^\alpha$, but more advanced approaches exist \cite{po16candeloro}.

Generalizing density, {\em i.e.} the number of present links divided by the total number of possible links, raises subtle questions. Indeed, it seems natural to replace the number of present links by the sum of all weights $\sum_{uv \in E} \weight(uv)$ like for strength, but several variants for the total sum of possible weights make sense. For instance, the litterature on rich clubs \cite{sr14alstott,prl08opsahl} considers that all present links may have the maximal weight, leading to $\sum_{uv \in E} \weight_{\max}$, or that all links may be present and have the maximal weight, leading to $\sum_{uv \in V\otimes V} \weight_{\max}$. In the special case where weights represent a level of certainty between $0$ and $1$ for link presence ($1$ if it is present for sure, $0$ if it is absent for sure), then the weighted density may be defined as $\frac{\sum_{uv \in E} \weight(uv)}{|V \otimes V|}$ \cite{mlg12zou}.

Various definitions of weighted clustering coefficients have been proposed, and \cite{pre07saramaki,ddns08antoniou,nc17wang} review many of them in details. The most classical one was proposed in \cite{pnas04barrat}:
$\cc(v) = \frac{1}{s(v)(d(v)-1)}\sum_{i,j\in N(v), ij\in E} \frac{\weight(vi)+\weight(vj)}{2}$. A general approach was also proposed in \cite{sn09opsahl}. Given a node $v$, it assigns a value to each triplet of distinct nodes $(i,v,j)$ such that $iv$ and $jv$ are in $E$ and to each such triplet such that $ij$ is also in $E$.
Then, $\cc(v)$ is defined as the ratio between the sum of values of triplets in the second category and the one of triplets in the first category.
In \cite{sn09opsahl}, considered values are the arithmetic mean, geometric mean, maximal value or minimal value of weights of involved links, depending on the application.
One may also consider the product of weights, leading to $\cc(v) = \frac{\sum_{i,j \in N(v), ij\in E} \weight(vi)\cdot \weight(vj) \cdot \weight(ij)}{\sum_{i\neq j \in N(v)} \weight(vi)\cdot \weight(vj)}$ as proposed in \cite{sagmb05zhang,ac07kalna,pre07ahnert} with normalized weights.

Transitivity is generalized in a very similar way \cite{sn09opsahl} by considering all triplets of distinct nodes $(i,j,k)$ such that $ij$ and $jk$ are in $E$ and each such triplet such that $ik$ is also in $E$, instead of only the ones centered on a specific node $v$.
If the associated  value is the product of weight, this leads to $\tr = \frac{\sum_{(i,j,k)} \weight(ij)\cdot \weight(jk) \cdot \weight(ik)}{\sum_{(i,j,k)} \weight(ij)\cdot \weight(jk)}$. If all weights are equal to $1$ ({\em i.e.} the graph is unweighted) this is nothing but the transitivity in $G$.

Various other concepts have been generalized to weighted graphs, like for instance assortativity \cite{pnas04barrat}, and specific weighted graph concepts, like closeness and betweenness centralities \cite{sn10opsahl}, connectability \cite{sr18amano}, eigenvector centrality \cite{pre04newman}, or rich club coefficient \cite{sr14alstott,prl08opsahl,epjb09zlatic}. We do not consider them here as our focus is on the most basic properties.

\medskip

\begin{figure}[!h]
\centering
\includegraphics[scale=.5]{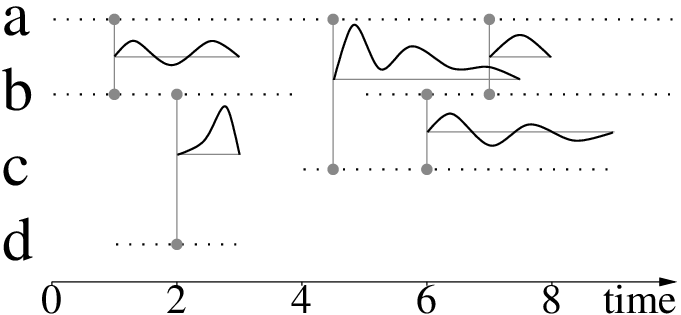}
\caption{
{\bf An example of weighted stream graph.} In this example, nodes are unweighted but links are weighted. Instead of just a straight horizontal line indicating link presence over time, we plot the weight value (assuming that $0$ is indicated by the horizontal line).
}
\label{fig:weighted}
\end{figure}

We define a {\bf weighted stream graph} as a stream graph $S = (T,V,W,E)$ equipped with a weight function $\weight$ defined over $W$ and $E$: if $(t,v) \in W$ then $\weight(t,v)$ is the weight of node $v$ at time $t$, and if $(t,uv) \in E$ then $\weight(t,uv)$ is the weight of link $uv$ at time $t$. See Figure~\ref{fig:weighted} for an illustration.

If a stream graph represents money transfers, then node weights may represend available credit and link weights may represent transfer amounts; if a stream represents travels, then node weights may represent available fuel, and link weights may represent speed; if a stream represents contacts between mobile device, node weights may represent battery charge and link weights may represent signal strength or link capacity; if a stream represents data transfers between computers then link weights may represent throughput or error rates; like for weighted graphs, countless situations may benefit from a weighted stream graph modeling.

As we will see in Section~\ref{sec:bipartite}, weighted stream graphs also widely appear within bipartite graph studies. In addition, as explained in \cite{stream}, section~19, one often resorts to $\Delta$-analysis for stream graph studies. Given a stream graph $S = (T,V,W,E)$ with $T=[x,y]$ and a parameter $\Delta$, it most simple form consists in transforming $S$ into $S_\Delta = (T',V,W', E')$ such that $T' = [x+\demidelta, y-\demidelta]$,
$T'_v = T' \cap \{t, \exists t' \in [t-\demidelta,t+\demidelta], t' \in T_v\}$, and
$T'_{uv} = T' \cap \{t, \exists t' \in [t-\demidelta,t+\demidelta], t' \in T_{uv}\}$.
Then, one may capture the amount of information in $S$ leading to node and link presences in $S_\Delta$ with weights: $\weight(t,v) = |\{t, \exists t' \in [t-\demidelta,t+\demidelta], t' \in T_v\}|$ and $\weight(t,uv) = |\{t, \exists t' \in [t-\demidelta,t+\demidelta], t' \in T_{uv}\}|$.

Like for weighted graphs, in addition to $S$ itself (without weights), one may consider the thresholded (unweighted) stream graphs $S_{\tau} = (T,V,W_\tau,E_\tau)$ where $W_\tau = \{(t,v) \in W, \weight(t,v) \ge \tau\}$ and $E_\tau = \{(t,uv) \in E, \weight(t,uv)\ge\tau\}$, for various thresholds $\tau$. One may then study how the properties of $S_{\tau}$ evolve with $\tau$.

The graph obtained from $S$ at time $t$, $G_t = (V_t,E_t)$, is naturally weighted by the function $\weight_t(v) = \weight(t,v)$ and $\weight_t(uv)=\weight(t,uv)$.
Likewise, the induced graph $G(S)$ is weighted by $\weight(v) = \frac{1}{|T|} \int_{t\in T_v} \weight(t,v) \diff t$ and $\weight(uv) = \frac{1}{|T|} \int_{t\in T_{uv}} \weight(t,uv) \diff t$. These definitions correspond to the average weight over time of each node and link. One may define similarly the minimal and maximal node and link weights, and go further by studying $S$ through the weighted graph $G(S)$ and the time-evolution of weighted graph properties of $G_t$.

These approaches aim to take both the weight and the temporal aspect into account:
from a weighted stream graph, the first one provides a family of (unweighted) stream graphs, one for each considered value of the threshold;
the second one provides a series of (static) weighted graphs, one for each instant considered.
In both cases, the actual combination of weight and time information is poorly captured.
We will therefore define concepts that jointly deal with both weights and time. Like with weighted graphs, we simplify the presentation by assuming that only links are weighted (nodes are not).

Since the degree of node $v$ in a stream graph $S$ is $d(v) = \sum_{u \in V} \frac{|T_{uv}|}{|T|}$ and since the strength of node $v$ in a weighted graph $G$ is $s(v) = \sum_{u\in N(v)} \weight(uv)$, we define the strength of node $v$ in a weighted stream graph $S$ as $s(v) = \sum_{u\in V}\int_{t\in T_{uv}} \frac{\weight(t,uv)}{|T|} \diff t$.
It is the degree of $v$ where each neighbor is counted with respect to the weight of its links with $v$ at the times when it is linked to $v$. It is related to the strength $s_t(v)$ of $v$ in $G_t$ as follows: $s(v) = \frac{1}{|T|} \int_{t \in T_v} s_t(v)$; it is the average strength of $v$ over time.

With this definition, one may study correlations between degree and strength in stream graphs, as with graphs, and even directly use their combinations, like $d(v)\cdot\left(\frac{s(v)}{d(v)}\right)^\alpha$ where $\alpha$ is a parameter.

Unsurprisingly, generalizing density to weighted stream graphs raises the same difficulties as for weighted graphs. Still, proposed definitions for weighted graphs easily apply to weighted stream graphs. Indeed, the sum of all weights becomes $\sum_{uv \in V\otimes V}\int_{t \in T_{uv}} \weight(t,uv) \diff t$, and the maximal weight of possible links may be defined as $\sum_{uv \in V\otimes V}\int_{t \in T_{uv}} \weight_{\max} \diff t = \weight_{\max} \cdot |E|$ or $\sum_{uv \in V\otimes V}\int_{t \in T} \weight_{\max} \diff t = \weight_{\max} \cdot |V \otimes V|$. Like with weighted graphs, if all weights are in $[0,1]$ then the weighted density may be defined by $\frac{\sum_{uv \in V\otimes V}\int_{t \in T_{uv}} \weight(t,uv) \diff t}{|T\otimes V\otimes V|}$.

Then, one may define the clustering coefficient of $v$ as the weighted density (according to one of the definitions above or another one) of the neighborhood of $v$. One may also consider the time-evolution of one the weighted clustering coefficient in $G_t$, according to previously proposed definitions surveyed above. Interestingly, one may also generalize the classical definition \cite{pnas04barrat} as follows:
$\cc(v) = \frac{1}{2s(v)(d(v)-1)} \int_{t\in T_v}\sum_{i,j \in N_t(v),(t,ij)\in E}\weight(t,vi)+\weight(t,vj)\diff t$, which is the product of link weights of $v$ with its pairs of neighbors when these neighors are linked together.

The general approach of \cite{sn09opsahl} also extends: one has to assign a value to each quadruplet $(t,i,j,k)$ with $i$, $j$, and $k$ distinct such that $(t,ij)$, $(t,jk)$ are in $E$, and to each such quadruplet such that $(t,ik)$ also is in $E$. 
As in the weighted graph case, the weighted stream graph clustering coefficient of node $v$, $\cc(v)$ is then the ratio between the sum of values of quadruplets in the second category such that $j=v$ and the one of quadruplets in the first category such that $j=v$ too. If the value of a quadruplet is the product of the weights of involved links, we obtain
$\cc(v) = \frac{\int_{t\in T_v}\sum_{i,j \in N_t(v), (t,ij)\in E} \weight(t,vi)\cdot \weight(t,vj) \cdot \weight(t,ij) \diff t}{\int_{t\in T_v}\sum_{i\neq j \in N_t(v)} \weight(t,vi)\cdot \weight(t,vj)}$.

Likewise, we define the weighted stream graph transitivity as the ratio between the sum of values of all quadruplets in the second category above and the one of quadruplets in the first category. If the value of quadruplets is defined as the product of weights of involved links, this leads to $\tr = \frac{\int_t \sum_{(i,j,k)} \weight(t,ij)\cdot \weight(t,jk) \cdot \weight(t,ik) \diff t}{\int_t \sum_{(i,j,k)} \weight(t,ij)\cdot \weight(t,jk) \diff t}$.
If all weights are equal to $1$ ({\em i.e.} the stream is unweighted), this is nothing but the stream graph transitivity defined in \cite{stream}.

\medskip

If $S$ is a graph-equivalent stream weighted by a constant function over time, {\em i.e.} $\weight(t,v)=\weight(t',v)$ and $\weight(t,uv) = \weight(t',uv)$ for all $t$ and $t'$, then it is equivalent to the weighted graph $G(S)$ weighted by $\weight(v) = \weight(t,v)$ and $\weight(uv) = \weight(t,uv)$ for any $t$; we call it a weighted graph-equivalent weighted stream.
The strength of $v$ in $S$ is equal to its strength in $G(S)$ if $S$ is a weighted graph-equivalent weighted stream.
The same is true for the different notions of density or clustering coefficient:
the density of $S$ is equal to the density of $G(S)$
and the clustering coefficient of a vertex $v$ or the transitivity are equal to the
clustering coefficient or transitivity in $G(S)$.


\section{Bipartite stream graphs}
\label{sec:bipartite}

A {\bf bipartite graph} $G=(\top\cup\bot,E)$ is defined by a set of top nodes $\top$, a set of bottom nodes $\bot$ with $\top \cap \bot = \emptyset$, and a set of links $E \subseteq \top \otimes \bot$: there are two different kinds of nodes and links may exist only between nodes of different kinds. Like weighted graphs, but maybe less known, bipartite graphs are pervasive and model many real-world data \cite{ipl04guillaume,sn08latapy}.
Typical examples include relations between client and products \cite{sigkdd14bernardes}, between company boards and their members \cite{robins04small,battistion03statistical}, and between items and their key features like movie-actor networks \cite{watts98collective,newman01random} or publication-author networks \cite{newman01scientific1,newman01scientific2}, to cite only a few.

Bipartite graphs are often studied through their top or bottom projections \cite{breiger74duality} $G_\top = (\top,E_\top)$ and $G_\bot = (\bot,E_\bot)$, defined by $E_\top = \cup_{v\in \bot} N(v) \otimes N(v)$ and $E_\bot = \cup_{v\in \top} N(v) \otimes N(v)$. In other words, in $G_\top$ two (top) nodes are linked together if they have (at least) a (bottom) neighbor in common in $G$, and $G_\bot$ is defined symmetrically. Notice that, if $v\in\top$ (resp. $v\in\bot$), then $N(v)$ always is a (not necessarily maximal) clique in $G_\bot$ (resp. $G_\top$).

Projections induce important information losses: the existence of a link or a clique in the projection may come from very different causes in the original bipartite graph. To improve this situation, one often considers {\em weighted} projections: each link $uv$ in the projection is weighted by the number $\weight(uv) = N(u) \cap N(v)$ of neighbors $u$ and $v$ have in common in the original bipartite graph. One may then use weighted graph tools to study the weighted projections \cite{iptps05,pnas04barrat,pre04newman}, but information losses remain important. In addition, projection are often much larger than the original bipartite graphs, which raises serious computational issues \cite{sn08latapy}.

As a consequence, many classical graph concepts have been extended to deal directly with the bipartite case, see \cite{sn08latapy,borgatti97network,faust97centrality,breiger74duality,bonacich72technique}. The most basic properties are $n_\top = |\top|$ and $n_\bot = |\bot|$, the number of top and bottom nodes. 
The definition of the number of links $m=|E|$ is the same as in classical graphs.
The definitions of node neighborhoods and degrees are also unchanged. The average top and bottom degrees $d_\top$ and $d_\bot$ of $G$ are the average degrees of top and bottom nodes, respectively.

With these notations, the bipartite density of $G$ is naturally defined as $\delta(G) = \frac{m}{n_\top \cdot n_\bot}$: it is the probability when one takes two nodes that may be linked together that they indeed are. Then, a bipartite clique in $G$ is a set
$C_\top \cup C_\bot$ with $C_\top \subseteq \top$ and $C_\bot \subseteq \bot$
such that $C_\top \times C_\bot \subseteq E$. In other words, all possible links between nodes in a bipartite clique are present in the bipartite graph.

Several bipartite generalizations of the clustering coefficient have been proposed \cite{sn08latapy,pre05lind,pa08zhang,sn13opsahl,ictir16lioma}. In particular:
\begin{itemize}
\item
  \cite{sn08latapy} and \cite{pre05lind} rely on the Jaccard coefficient defined over node pairs
  both either in $\top$ or $\bot$:
  $\cc(uv) = \frac{|N(u)\cap N(v)|}{|N(u)\cup N(v)|}$ or close variants.
One then obtains the clustering coefficient of node $v$ by averaging its Jaccard coefficient with all neighbors of its neighbors: $\cc(v) = \frac{\sum_{u\in N(N(v)), u\ne v} \cc(uv)}{|N(N(v))|}$.
\item
for each node $v$, \cite{sn08latapy} and \cite{ictir16lioma} consider all triplets $(u,v,w)$ of distinct nodes with $u$ and $w$ in $N(v)$ and define the redundancy $\rc(v)$ as the fraction of such triplets such that there exists an other node in $N(u)\cap N(w)$.
\item
similarly, \cite{sn13opsahl} proposes to consider all quintuplets $(a,b,v,c,d)$ of distinct nodes with $b, c\in N(v)$, $a\in N(b)$ and $d\in N(c)$ and defines $cc^*(v)$ as the fraction of them such that there exists another node in $N(a)\cap N(d)$.
\end{itemize}

Transitivity is usually extended to bipartite graphs \cite{robins04small,sn08latapy,sn13opsahl} by considering the set ${\sf N}$ of quadruplets of nodes $(a,b,c,d)$ such that $ab$, $bc$ and $cd$ are in $E$ and the set $\bowtie$ of such quadruplets with in addition $ad$ in $E$. Then, the transitivity is $\tr(G) = \frac{|\bowtie|}{|{\sf N}|}$.
Like above, \cite{sn13opsahl} also propose to consider the fraction of all quintuplets $(a,b,c,d,e)$ of distinct nodes with $ab$, $bc$, $cd$, $de$ in $E$ such that there exists an other node $f$ with $af$ and $ef$ in $E$.

\medskip

\begin{figure}[!h]
\centering
\ \hfill
\includegraphics[scale=.5]{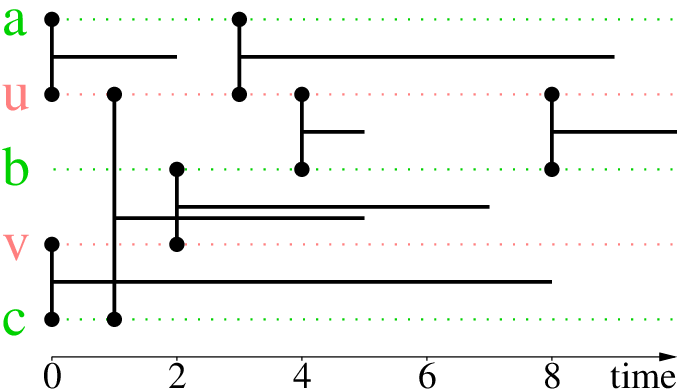}
\hfill
\includegraphics[scale=.5]{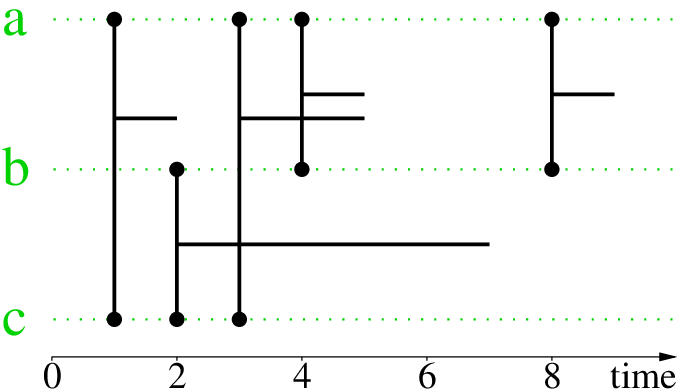}
\hfill \ 
\caption{
{\bf Left:} a bipartite stream graph $S = (T,\top\cup\bot,W,E)$ with $T=[0,10]$, $\top=\{u,v\}$, $\bot=\{a,b,c\}$, $W = T \times (\top\cup\bot)$, and $E = ([0,2]\cup[3,9])\times\{ua\} \cup ([4,5]\cup[8,10])\times\{ub\} \cup [1,5]\times\{uc\} \cup [2,7]\times\{v,b\} \cup [0,8]\times\{vc\}$.
{\bf Right:} its $\bot$-projection $S_\bot$. For instance, $a$ and $c$ are linked together from time $3$ to $5$ because they both have $u$ in their neighborhood for this time period in $S$.
}
\label{fig:bipartite}
\end{figure}

We define a {\bf bipartite stream graph} $S = (T,\top\cup\bot,W,E)$ from a time interval $T$, a set of top nodes $\top$, a set of bottom nodes $\bot$ with $\top\cap\bot=\emptyset$, and two sets $W \subseteq T \times (\top\cup\bot)$ and $E \subseteq T \times \top \otimes \bot$ such that $(t,uv) \in E$ implies $(t,u) \in W$ and $(t,v) \in W$. See Figure~\ref{fig:bipartite} (left) for an illustration.
Each instantaneous graph $G_t$, as well as the induced graph $G(S)$, are bipartite graphs with the same top and bottom nodes.

Bipartite stream graphs naturally model many situations,
like for instance presence of people in rooms or other kinds of locations,
purchases of products by clients, access to on-line services,
bus presence at stations \cite{mpe19curzel}, traffic between a set of computers and the rest of the internet \cite{cn18viard},
or contribution of people to projects, such as software.

The classical definition of projections is easily extended, leading to $S_\top = (T,\top,W_\top,E_\top)$ and $S_\bot = (T,\bot,W_\bot,E_\bot)$, where $W_\top = W \cap (T\times\top)$, $W_\bot = W \cap (T\times\bot)$, $E_\top = \cup_{(t,v)\in W_\bot} \{(t,uw) \mbox{ s.t. } (t,uv)\in E \mbox{ and } (t,wv)\in E\}$ and $E_\bot = \cup_{(t,v)\in W_\top} \{(t,uw) \mbox{ s.t. } (t,uv)\in E \mbox{ and } (t,vw)\in E\}$. In other words, in $S_\top$ two (top) nodes are linked together at a given time instant if they have (at least) a (bottom) neighbor in common in $S$ at this time, and $S_\bot$ is defined symmetrically. See Figure~\ref{fig:bipartite} for an illustration. Notice that, if $v\in\top$ (resp. $v\in\bot$) then $N(v)$ always is a (not necessarily maximal) clique in $S_\bot$ (resp. $S_\bot$).

One may also generalize weighted projections by considering the number
$\weight(t,uv) = |N_t(u) \cap N_t(v)|$ of neighbors $u$ and $v$ have in common at time $t$
in the original bipartite stream graph. One then obtains weighted stream graphs, and may use the definitions proposed in Section~\ref{sec:weighted} to study them. Still, this induces much information loss, which calls for the generalization of bipartite properties themselves.

The most immediate definitions are the numbers of top and bottom nodes: $n_\top = \sum_{v\in \top} \frac{|T_v|}{|T|} = \frac{|W \cap (T\times\top)|}{|T|}$ and $n_\bot = \sum_{v\in \bot} \frac{|T_v|}{|T|} = \frac{|W \cap (T\times\bot)|}{|T|}$, respectively. We then have $n = n_\top + n_\bot$. Like with graphs, the number of links, neighborhood of nodes, and their degree do not call for specific bipartite definitions.
We however define the average top and bottom degrees $d_\top$ and $d_\bot$ of $S$ as the average degrees of top and bottom nodes, respectively, weighted by their presence time in the stream:
$d_\top(S) = \sum_{v\in \top} \frac{|T_v|}{|W|} d(v)$ and $d_\bot(S) = \sum_{v\in \bot} \frac{|T_v|}{|W|} d(v)$.


We define the bipartite density of $S$ as
$\delta(S) = \frac{m}{\sum_{u\in\top,v\in\bot}|T_u\cap T_v|}$:
it is the probability when one takes two nodes when they may be linked together that they indeed are.
A subset $C=C_\top \cup C_\bot$ of $W$, with $C_\top \subseteq T\times\top$ and $C_\bot \subseteq T\times\bot$, is a clique in $S$ if all possible links exist between nodes when they are involved in $C$: for all $t$, if $(t,u)$ is in $C_\top$ and $v$ is in $C_\bot$ then $uv$ is in $E$.

Like with graphs, defining a bipartite stream graph clustering coefficient is difficult and leaves us with several reasonable choices:
\begin{itemize}
\item
  Extending the Jaccard coefficient to node pairs in a stream graph leads to an instantaneous definition: $cc_t(uv) = \frac{|N_t(u)\cap N_t(v)|}{|N_t(u)\cup N_t(v)|}$, which is nothing but $\cc(uv)$ in $G_t$. It also leads to a global definition: $\cc(uv) = \frac{|N(u)\cap N(v)|}{|N(u)\cup N(v)|} = \frac{\sum_{w\in\top\cup\bot}|T_{uw}\cap T_{vw}|}{\sum_{w\in\top\cup\bot}|T_{uw}\cup T_{vw}|}$.
  We may then define the bipartite clustering coefficient of node $v$ by averaging its Jaccard coefficient with all neighbors of its neighbors, weighted by their co-presence time:
$\frac{1}{|N(N(v))|}\sum_{u\in \top\cup\bot} \frac{|T_u \cap T_v|}{|T|}\cc(uv)$.
Notice that $\cc(uv) = 0$ if 
the neighborhoods of $u$ and $v$ do not intersect.
This means that the sum actually is over nodes that are at some time neighbor of a neighbor of $v$, which is consistent with the bipartite graph definition.
\item
Redundancy is easier to generalize to stream graphs:
for each node $v$, we consider all triplets $(t,u,v,w)$ composed of a time instant $t$, a neighbor $u$ of $v$ at time $t$, $v$ itself, and another neighbor $w$ of $v$ at time $t$, and we define $\rc(v)$ as the fraction of such quadruplets such that there exists an other node linked to $u$ and $w$ at time $t$. In other words,
$\rc(v) = \frac{|\{(t,u,v,w), u\neq w, u, w\in N_t(v), \exists x\in N_t(u)\cap N_t(w), x\ne v\}|}{|\{(t,u,v,w), u\neq w, u, w\in N_t(v)\}|}$.

\item
Similarly, we propose a stream graph generalization of $cc^*(v)$ as the fraction of sextuplets $(t,a,b,v,c,d)$ with $a$, $b$, $v$, $c$, and $d$ all different from each other, $b\in N_t(v)$, $c\in N_t(v)$, $a\in N_t(b)$ and $d\in N_t(c)$ for which in addition there exists an other node in $N_t(a)\cap N_t(d)$.
\end{itemize}

Finally, if we denote by ${\sf N}$ the set of quintuplets $(t,a,b,c,d)$ such that $(t,ab)$, $(t,bc)$ and $(t,cd)$ are in $E$ and the set $\bowtie$ of such quintuplets with in addition $(t,ad)$ in $E$, then the bipartite transitvity for stream graphs is $\tr(G) = \frac{|\bowtie|}{|{\sf N}|}$ as before. Like for bipartite graphs, one may also consider the fraction of all sextuplets $(t,a,b,c,d,e)$ where $a$, $b$, $c$, $d$, and $e$ are distinct nodes with $(t,ab)$, $(t,bc)$, $(t,cd)$, $(t,de)$ in $E$ such that there exists an other node $f$ with $(t,af)$ and $(t,ef)$ in $E$.

\medskip

If $S$ is a graph-equivalent bipartite stream, then its projections $S_\top$ and $S_\bot$ are also graph-equivalent streams, and their corresponding graphs are the projections of the graph corresponding to $S$: $G(S_\bot) = G(S)_\bot$ and $G(S_\top) = G(S)_\top$. In addition, the bipartite properties of $S$ are equivalent to the bipartite properties of its corresponding bipartite graph:
the density, Jaccard coefficient, redundancy, and $cc^*$, as well as transitivity values,
are all equal to their graph counterpart in $G(S)$ if $S$ is
a graph-equivalent bipartite stream.

\section{Directed stream graphs}
\label{sec:directed}

A {\bf directed graph} $G=(V,E)$ is defined by its set $V$ of nodes and its set $E \subseteq V\times V$ of links: while links of undirected graphs are unordered pairs of distinct nodes, links in directed graphs are ordered pairs of nodes, not necessarily distinct: $(u,v) \neq (v,u)$, and $(v,v)$ is allowed and called a loop. Then, $(u,v)$ is a link from $u$ to $v$, and if both $(u,v)$ and $(v,u)$ are in $E$ then the link is said to be symmetric.

Directed graphs naturally model the many situations where link asymmetry is important, like for instance dependencies between companies or species, citations between papers or web pages \cite{jws15meusel}, friendship relations in many on-line social networks \cite{imc07mislove}, or hierarchical relations of various kinds.

Directed graph are often studied as undirected graphs by ignoring link directions. However, this is not satisfactory in many cases: having a link to a node is very different from having a link from a node and, for instance, having links to many nodes is very different from having links from many nodes. As a consequence, many directed extensions of standard graph properties have been proposed to take direction into account, see for instance \cite{wasserman1994social} section~4.3 and \cite{pre07fagiolo,jfi65hakimi}.

First, a node $v$ in a directed graph has two neighborhoods: its out-going neighborhood $N^+(v) = \{u, (v,u)\in E\}$ and its in-coming neighborhood $N^-(v) = \{u, (u,v)\in E\}$. This leads to its out- and in-degrees $d^+(v) = |N^+(v)|$ and $d^-(v) = |N^-(v)|$. These definitions make it possible to study the role of nodes with high in- and out-degrees, as well as correlations between these metrics \cite{imc07mislove,iwdc04}. Notice that $\sum_{v \in V} d^+(v) = \sum_{v \in V} d^-(v) = m$ is the total number of links, and so the average in- and out-degrees are equal to $\frac{m}{n}$.

The directed density of $G$ is $\frac{m}{n^2}$ since in a directed graph (with loops) there are $n^2$ possible links. Then, a directed clique is nothing but a set of nodes all pairwise linked together with symmetrical links, which, except for loops, is equivalent to an undirected clique. One may also consider the fraction of loops present in the graph $\frac{|\{ (v,v) \in E\}|}{n}$, as well as the fraction of symmetric links $\frac{|\{(u,v)\in E \mbox{ s.t. } (v,u)\in E\}|}{m}$.

With this definition of density, one may define the in- and out-clustering coefficient of node $v$ as the density of its in-coming and out-going neighborhood. However, this misses the diversity of ways $v$ may be linked to its neighbors and these neighbors may be linked together \cite{pre07fagiolo,wasserman1994social}. Table~1 in \cite{pre07fagiolo} summarizes all possibilities and corresponding extensions of clustering coefficient. We focus here on two of these cases, that received more attention because they capture the presence of small cycles and the transitivity of relations. Given a node $v$, the cyclic clustering coefficient is
the fraction of its pairs of distinct neighbors $u$ and $w$ with $(u,v)$ and $(v,w)$ in $E$,
such that $(w,u)$ also is in $E$, {\em i.e.} $u$, $v$ and $w$ form a cycle. The transitive coefficient of $v$ consists in the fraction of its pairs of distinct neighbors $u$ and $w$ with $(u,v)$ and $(v,w)$  in $E$, such that $(u,w)$ also is in $E$, {\em i.e.} the relation is transitive.

Finally, these extensions of node clustering coefficient directly translate to extensions of graph transitivity ratio. In the two cases explained above, this leads to the fraction of all triplets of distinct nodes $(u,v,w)$ with $(u,v)$ and $(v,w)$ in $E$, such that in addition $(w,v)$ is in $E$, or $(v,w)$ is in $E$, respectively.

\medskip

\begin{figure}[!h]
\centering
\includegraphics[scale=.5]{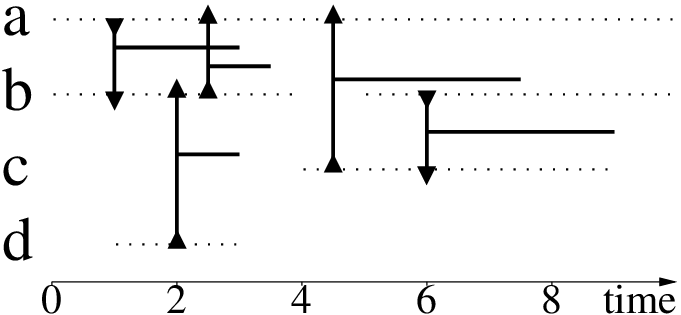}
\caption{
{\bf An example of directed stream graph}
$S=(T,V,W,E)$ with $T = [0,10] \subseteq \mathbb{R}$, $V = \{a,b,c,d\}$, $W = [0,10]\times\{a\} \cup ([0,4]\cup[5,10])\times\{b\} \cup [4,9]\times\{c\} \cup [1,3]\times\{d\}$, and $E = [1,3]\times\{(a,b)\} \cup [2.5,3.5]\times\{(b,a)\} \cup [4.5,7.5]\times\{(c,a)\} \cup [6,9]\times\{(b,c)\} \cup [2,3]\times\{(d,b)\}$.
Notice that $T_{a,b} = [1,3] \ne T_{b,a} = [2.5,3.5]$, and that links between $a$ and $b$ are symmetrical from time $2.5$ to time $3$.
}
\label{fig:directed}
\end{figure}

We define a {\bf directed stream graph} $S=(T,V,W,E)$ from a time interval $T$, a set of nodes $V$, a set of temporal nodes $W \subseteq T\times V$, and a set of temporal links $E \subseteq T\times V\times V$: $(t,u,v)$ in $E$ means that there is a link from $u$ to $v$ at time $t$, which is different from $(t,v,u)$. We therefore make a distinction between $T_{u,v}$ the set $\{t, (t,u,v)\in E\}$ and $T_{v,u}$ the set $\{t, (t,v,u)\in E\}$. A loop is a triplet $(t,v,v)$ in $E$. If both $(t,u,v)$ and $(t,v,u)$ are in $E$, then we say that this temporal link is symmetric.
See Figure~\ref{fig:directed} for an illustration.

Directed stream graphs model the many situations where directed links occur over time and their asymmetry is important, like for instance money transfers, network traffic, phone calls, flights, moves from a place to another, and many others. In all theses cases, both time information and link direction are crucial, as well as their interplay. For instance, a large number of computers sending packets at a given computer in a very short period of time is a typical signature of denial of service attack \cite{ijnm15mazel}. Instead, a computer sending packets to a large number of other computers in a short period of time is typical of a streaming server.

The directed stream graph $S = (T,V,W,E)$ may be studied through the standard stream graph $(T,V,W,\{(t,uv), (t,u,v)\in E \mbox{ or } (t,v,u) \in E\})$ obtained by considering each directed link as undirected. Likewise, $S$ may be studied through its induced directed graph $G(S)=(\{v, \exists (t,v)\in W\},\{(u,v), \exists (t,u,v)\in E\})$ and/or the sequence of its instantaneous directed graphs $G_t = (\{v, (t,v) \in W\},\{(u,v), (t,u,v)\in E\})$. However, these approaches induce much information loss, and make it impossible to make subtle distinctions like the one described above for network traffic. This calls for generalizations of available concepts to this richer case.

We define the out-going neighborhood of node $v$ as $N^+(v) = \{(t,u), (t,v,u) \in E\}$ and its in-coming neighborhood as $N^-(v) = \{(t,u), (t,u,v) \in E\}$.
Its in- and out-degrees are $d^+(v) = \frac{|N^+(v)|}{|T|} = \sum_{u\in V} \frac{|T_{v,u}|}{|T|}$ 
and $d^-(v) = \frac{|N^-(v)|}{|T|} = \sum_{u\in V} \frac{|T_{u,v}|}{|T|}$. 
Like with directed graphs, the total number of links $m$ is equal to $\sum_{v \in V} d^+(v)$ as well as to $\sum_{v \in V} d^-(v)$.

We extend the standard stream graph density 
$\frac{m}{\sum_{uv \in V\otimes V}|T_u \cap T_v|}$ 
into the directed stream graph density $\frac{m}{\sum_{(u,v) \in V\times V}|T_u \cap T_v|}$: it is the fraction of possible links that indeed exist. 
Then, a clique in a directed stream graph is a subset $C$ of $W$ such that for all $(t,u)$ and $(t,v)$ in $C$, both $(t,u,v)$ and $(t,v,u)$ are in $E$. These definitions are immediate extensions of standard stream graph concepts.

One may then define the in- and out-clustering coefficient of a node $v$ as the density of its in- and out-neighborhoods in the directed stream graph. However, like for directed graphs, there are many possible kinds of interactions between neighors of a node, which lead to various generalizations of clustering coefficient to directed stream graph. We illustrate this on the two definitions detailed above for a given node $v$. First, let us consider the set of quadruplets $(t,u,v,w)$ with $u$, $v$ and $w$ distinct, such that $(t,u,v)$ and $(t,v,w)$ are in $E$. Then one may measure the cyclic clustering coefficient as the fraction of these quadruplets such that in addition $(t,w,u)$ is in $E$, and the transitive clustering coefficient as the fraction of these quadruplets such that in addition $(t,w,u)$ is in $E$.

Like with directed graphs, this leads to directed stream graph extensions of transitivity. In the two cases above, we define it as the fraction of quadruplets $(t,u,v,w)$ with $u$, $v$ and $w$ distinct and with $(t,u,v)$ and $(t,v,w)$ in $E$, such that in addition $(t,w,v)$ is in $E$, or $(t,v,w)$ is in $E$, respectively.

\medskip

If $S$ is a directed graph-equivalent directed stream graph, {\em i.e.} $G_t = G(S)$ for all $t$, then all the properties of $S$ defined above are equal to their directed graph counterpart in $G(S)$.

\section{Conclusion}
\label{sec:conclusion}


Previous works extended many graph concepts to deal with weighted, bipartite or directed graphs. In addition, graphs were generalized recently to stream graphs in order to model temporal networks in a way consistent with graph theory. In this contribution, we show that weighted, bipartite or directed graphs concepts may themselves be generalized to weighted, bipartite or directed stream graphs, in a way consistent with both their graph counterparts and the stream graph formalism. This opens the way to a much richer modeling of temporal networks, and more precise case studies, in a unified framework.

Such case studies may benefit from improved modeling with either weights, different sorts of nodes, or directed links,
but may also combine these extensions together. For instance, money transfers between clients and sellers are best modeled by weighted bipartite directed stream graphs.
The concepts we discussed then have to be extended even further, like what has already been done for graphs
for instance with the directed strength $s^+(v) = \sum_{u\in N^+(v)} \weight(u,v)$ \cite{sn10opsahl},
the directed weighted clustering coefficient~\cite{csf18clemente,pre07fagiolo},
or for the study of weighted bipartite graphs~\cite{iwdc04}.

One may also consider other kinds of graph extensions, like multigraphs, labelled graphs, hypergraphs, or multi-layer graphs for instance, which model important features of real-world data and already received much attention. We focused on weighted, bipartite and directed cases because they seem to be the most frequent in applications.



Likewise, we selected only a few key weighted, bipartite or directed properties in order to extend them to stream graphs. Many others remain to generalize, in particular path-related concepts like reachability, closeness, or betweenness, to cite only a few \cite{sn10opsahl}.

\medskip
\noindent
{\bf Acknowledgements.}
This work is funded in part by the European Commission
H2020 FETPROACT 2016-2017 program under grant 732942 (ODYCCEUS),
by the ANR (French National Agency of Research) under grants ANR-15-CE38-0001 (AlgoDiv),
by the Ile-de-France Region and its program FUI21 under grant 16010629 (iTRAC).

\bibliographystyle{plain}
\bibliography{biblio}

\end{document}